\documentclass[sigconf]{acmart}
\usepackage{dirtytalk}
\usepackage{tcolorbox}
\usepackage{tikz}
\usepackage{framed}
\usepackage{color}

\definecolor{formalshade}{rgb}{0.97,0.97,0.97}
\definecolor{darkblue}{rgb}{0.90,0.90,0.90}

\usepackage{fontawesome}

\AtBeginDocument{%
  \providecommand\BibTeX{{%
    \normalfont B\kern-0.5em{\scshape i\kern-0.25em b}\kern-0.8em\TeX}}}

\setcopyright{acmcopyright}
\copyrightyear{2022}
\acmYear{2022}
\acmDOI{AAAAAAAAA}

\acmConference[EASE '22]{EASE '22: International Conference on Evaluation and Assessment in Software Engineering}{June 13--15, 2022}{Göteborg, Sweden}
\acmBooktitle{EASE '22: International Conference on Evaluation and Assessment in Software Engineering,
  June 13--15, 2022, Göteborg, Sweden}
\acmPrice{AAAA}
\acmISBN{AAAAAAAAA}

\newcommand{\humphrey}[1]{}
\renewcommand{\humphrey}[1]{{\color{red} Humphrey: {#1}}}   
\newcommand{\Mojtaba}[2]{}
\renewcommand{\Mojtaba}[2]{{\color{blue} Mojtaba: {#2}}} 
\begin{document}
%
\title{Human Values Violations in Stack Overflow: An Exploratory Study}

 \author{Sara Krishtul$^{1}$, Mojtaba Shahin$^{2}$, Humphrey O. Obie$^{1}$\\Hourieh Khalajzadeh$^{1}$, Fan Gai$^{1}$, Ali Rezaei Nasab$^{3}$, John Grundy$^{1}$}
 \affiliation{%
   \institution{$^{1}$Faculty of IT, Monash University, Melbourne, Australia}
   \institution{$^{2}$School of Computing Technologies, RMIT University, Melbourne, Australia} 
   \institution{$^{3}$School of Electrical and Computer Engineering, Shiraz University, Shiraz, Iran}
  \institution{\{skri0002, fgai0001\}@student.monash.edu, mojtaba.shahin@rmit.edu.au, \{humphrey.obie, hourieh.khalajzadeh, john.grundy\}@monash.edu, alirezaei@hafez.shirazu.ac.ir}
  \country{}}

\renewcommand{\shortauthors}{Krishtul et al.}


\begin{abstract}
A growing number of software-intensive systems are being accused of \textit{violating or ignoring human values} (e.g., privacy, inclusion, and social responsibility), and this poses great difficulties to individuals and society. Such violations often occur due to the solutions employed and decisions made by developers of such systems that are misaligned with user values. Stack Overflow is the most popular Q\&A website among developers to share their issues, solutions (e.g., code snippets), and decisions during software development. We conducted an exploratory study to investigate the occurrence of human values violations in Stack Overflow posts. As comments under posts are often used to point out the possible issues and weaknesses of the posts, we analyzed 2,000 Stack Overflow comments and their corresponding posts (1,980 unique questions or answers) to identify the types of human values violations and the reactions of Stack Overflow users to such violations. Our study finds that 315 out of 2,000 comments contain concerns indicating their associated posts (313 unique posts) violate human values. Leveraging Schwartz’s theory of basic human values as the most widely used values model, we show that \textit{hedonism} and \textit{benevolence} are the most violated value categories. We also find the reaction of Stack Overflow commenters to perceived human values violations is very quick, yet the majority of posts (76.35\%) accused of human values violation do not get downvoted at all. Finally, we find that the original posters rarely react to the concerns of potential human values violations by editing their posts. At the same time, they usually are receptive when responding to these comments in follow-up comments of their own.
\end{abstract}



%

\maketitle


\section{Introduction} \label{sec:intro}
\emph{``Every line of code represents an ethical and moral decision; every bit of data collected, analyzed, and visualized has moral implications.''} Grady Booch \cite{booch2014human}
\vspace{4pt}

Over the last few years, the media has reported many examples of software-intensive systems that intentionally or unintentionally ignored or violated ethical and human values \cite{galhotra2017fairness, incidentDB}. These systems have sometimes posed irreversible damages and challenges to end-users (e.g., loss of life), society (e.g., ignoring or being biased against a particular gender), and the software industry (e.g., damaging the software creator's reputation). For example, Amazon's  ``Prime same-day delivery service'' was developed to provide all US citizens an equal and fair shopping experience \cite{Amazon_2016}. However, it has been found that it prevents black neighborhoods from receiving such a shopping experience. In another recent example, the Facebook AI-based feature recommendation system mistakenly labeled a video of Black men as `Primates' \cite{facebook_2021,incidentDB}.  

Human values such as privacy, inclusion, power, fairness, and pleasure are defined as something that is deemed important for an individual, a group of people, or a society \cite{schwartz1992universals}. Many definitions and models have been proposed for human values in social science \cite{cheng2010developing}. However, the most well-known and widely used one is Schwartz's theory of basic human values \cite{schwartz1992universals,schwartz2012overview}. Schwartz's theory of basic values includes ten universal values (\textit{self-direction}, \textit{stimulation}, \textit{hedonism}, \textit{achievement}, \textit{power}, \textit{security}, \textit{conformity}, \textit{tradition}, \textit{benevolence}, \textit{universalism}), which were identified through a survey of participants across 80 countries. While ethics are referred to as moral expectations that all individuals in a society agree upon, each person's values may differ from those of another \cite{whittle2021case, fieser2016ethics}. Hence, human values are \textit{``enduring beliefs that a specific mode of conduct or end state of existence is personally or socially preferable to an opposite or converse mode of conduct or end state of existence''} \cite{rokeach1973nature}.

Human values have extensively been studied in the human-computer interaction field since the late 1980s \cite{whittle2021case}. However, the software engineering community has recently attempted to develop new software engineering practices and techniques or adapt the existing ones (e.g., the \textit{value-based requirements engineering }method \cite{thew2018value} and the \textit{fairness-aware programming} technique \cite{albarghouthi2019fairness}) to operationalize human values in software (e.g., \cite{perera2020continual, ferrario2016values, harbers2015embedding}). Operationalizing human values in software is defined as \textit{``the process of identifying human values and translating them to accessible and concrete concepts so that they can be implemented, validated, verified, and measured in software''} \cite{shahin2021operationalizing}. The ultimate goal is to develop software systems that better reflect and respect human values. Other lines of research aimed to understand which types of human values are discussed by developers in GitHub's issue tracking systems \cite{nurwidyantoro2021human} and analyzed app reviews to determine which human values are ignored or violated by apps \cite{obie2021first, shams2020society}.

\begin{figure}
    \centering
    \includegraphics[width=0.85\linewidth]{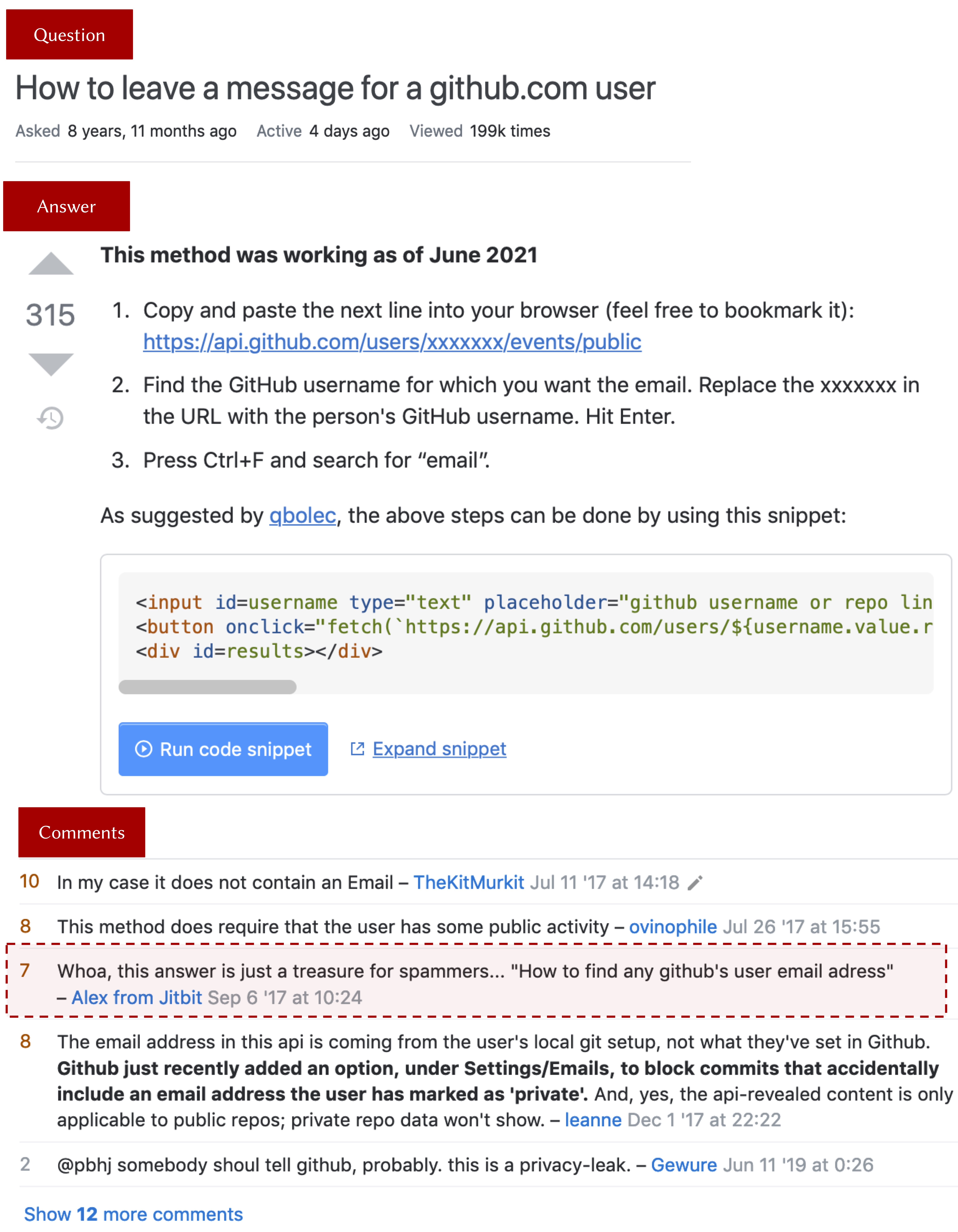}
    \caption{A question (ID: 12686545) with one of its answers and some of the comments under the answer. One comment criticizes a proposed answer as spammers can use it, e.g., for unethical purposes.}
    \label{fig:motivationexp}
\end{figure}

Inspired by \cite{booch2014human, whittle2021case, gotterbarn2018acm}, arguing that the codes developed or decisions made by developers may have a moral and ethical implication or a values implication, the objective of this study is to understand the human values-sensitive implications of developers’ solutions. Given Stack Overflow is the most active online platform for developers to share their programming issues and programming solutions (e.g., a code snippet or a design pattern as a solution), we focus on Stack Overflow. Each Stack Overflow post includes a question and a list of answers posted to the question \cite{zhang2019reading}. Stack Overflow provides mechanisms for users to comment on both questions and answers. This enables them to have further discussions (e.g., discuss the weaknesses of a solution) on the posted questions and answers. 

In this study, we analyzed 2000 comments and their associated questions or answers (i.e., 1980 unique questions or answers). We first studied which types of human values in Schwartz's theory of basic values are violated by Stack Overflow posts by manually analyzing comments. Then, we investigated the reactions of Stack Overflow users to human value violations. Key contributions of the work include:
\begin{itemize}
    \item To our knowledge, the first detailed study of human values violations in Stack Overflow posts
    \item A large number of Stack Overflow comments are manually analyzed against the Schwartz human values framework
    \item Identification of 315 comments using the framework that raise concerns that their 313 unique associated posts violate human values
    \item A set of recommendations for practitioners and researchers to address human value violations in SO posts
\end{itemize}

We first introduce our research motivation and questions in Section \ref{sec:reseachquestions}. Section \ref{sec:backgroundrelatedwork} provides the background and summarizes the related studies. In Section \ref{sec:datacollection}, we describe our data collection process, followed by reporting our findings in Section \ref{sec:findings}. We reflect on our findings in Section \ref{sec:discussion}. Possible threats of our study are reported in Section \ref{sec:threatstovalidity}. We conclude our paper in Section \ref{sec:conclusions}. 

\section{Motivation and Research Questions} \label{sec:reseachquestions}

Posts (questions and answers) in Stack Overflow may come with a number of issues. For example, a proposed solution in an answer might be obsolete \cite{zhang2019empirical} or even incorrect \cite{zhang2019reading}. Stack Overflow provides different mechanisms such as downvoting \cite{stackoverflow_downvoting} and commenting \cite{stackoverflow_comment} to enable Stack Overflow users to indicate the possible issues associated with posts. Stack Overflow recommends that question and answer comments be used to provide constructive criticism, to request clarification for a question or an answer from the poster, or to add relevant information \cite{stackoverflow_comment}. 
A question owner can also comment under their question, and comments under an answer can be added by the answer owner and the owner of its associated question. Any user with at least 50 reputation points can post comments under any question and answer \cite{stackoverflow_comment,zhang2019reading}. 

Previous research has shown that more than 50\% of both hidden and displayed comments are informative and can enhance their associated answers \cite{zhang2019reading}. Zhang et al. \cite{zhang2019reading} classified comments in Stack Overflow in seven categories, in which comments falling in \textit{advantage}, \textit{improvement}, \textit{weakness}, \textit{inquiry}, and \textit{addition} categories are considered informative. Hence, we argue that comments (in particular \textit{improvement}, \textit{weakness}, and \textit{inquiry} comments) can be the ideal place to point out problems and weaknesses of a post from a human values perspective. The reason behind this is that these types of comments try to challenge a question or answer.

Figure \ref{fig:motivationexp} shows a question (\textsc{\textbf{Question ID: 12686545}}) in Stack Overflow with one of its highly voted answers and some of the comments posted under the answer. As shown in Figure \ref{fig:motivationexp}, a user posted a comment to criticize the proposed answer because spammers can use it, for example, to send thousands of unsolicited emails to GitHub's users resulting in upsetting or annoying them. Hence, we argue that this answer has a values implication (e.g., violating the value of \textit{hedonism} - value item of \textit{pleasure}). We developed the following research questions that we wanted to answer in this study:

\textbf{RQ1.} \textit{Which types of human values are
perceived to be violated in Stack Overflow posts?}

\noindent\textbf{\underline{Motivation}}. As discussed, comments under a post can be used by Stack Overflow users to point out problems and weaknesses of a post from different perspectives, such as a human values perspective. This research question aims to identify types of values that are violated in Stack Overflow posts, and the frequency with which each value type is violated, by analyzing the claims made in Stack Overflow comments. This question can provide deep insights for Stack Overflow askers and answerers on potential human values-implications of their posts and the Stack Overflow community to develop mechanisms to recognize such posts.



\textbf{RQ2.} \textit{How quick are commenters to raise concerns over human values violations?}

\noindent\textbf{\underline{Motivation}}. A post (question or answer) might get several comments. Comments under a post are sorted and displayed by their creation time. As discussed earlier, the nature of comments varies from praising a post to criticizing and pointing out possible issues. This research question aims to understand how quickly comments citing human values violations are added to a post and determine their position among all comments under a post. 

\textbf{RQ3.} \textit{Are posts accused of violating human values downvoted by Stack Overflow users?}

\noindent\textbf{\underline{Motivation}}. Apart from the commenting mechanism, the downvoting mechanism in Stack Overflow also enables the community to indicate the posts (questions and answers) with deficiencies, e.g., incorrect answers \cite{stackoverflow_downvoting}. While voting down questions is free, voting down answers comes with some costs \cite{stackoverflow_downvoting}. A limited number of downvotes can be made by a user (voter) per day. Voting down an answer diminishes two reputations from the post owner and one reputation from the voter. Hence, voting down an answer should be done carefully. In this research question, we want to know whether Stack Overflow users vote down posts that violate human values despite the associated costs.


\textbf{RQ4.} \textit{How does the original poster react to concerns of potential human values violations?}

\noindent\textbf{\underline{Motivation}}. Once human values violations occur in a software system, its creators are expected to react to such violations adequately (e.g., fixing values violations immediately). In this research question, we aim to understand the reactions of Stack Overflow posters (askers and answerers) once issues of potential human values violations are raised concerning their posts. Such reactions can range from completely denying the violation to modifying the post to mitigate the violation.

\section{Background and Related Work} \label{sec:backgroundrelatedwork}
\subsection{Human Values in Software Engineering}

Human values are the guiding principles for what people consider important in life \cite{cheng2010developing}. Although these principles are often unarticulated using formal terminologies, they undergird people's decisions, technologists and non-technical people alike. Hence, the influence of human values can be detected in people’s preferences, from the choice of end-user applications \cite{obie2021does}, to the technical design decisions of developers in software engineering projects \cite{Winter:2019}.

The study of human values in software engineering (SE) is often based on Schwartz’s theory of basic human values \cite{perera2020study}. This theory organizes values into 10 broad values and is established on surveys conducted in multiple countries covering a wide range of ages, genders, occupations, cultural backgrounds, and geography \cite{schwartz1992universals}. Table \ref{tab:valuescategories} shows the 10 value categories. The 10 value categories comprise 58 value items, e.g., the value category of \textit{\textbf{benevolence}} is comprised of the value items of \textit{\textbf{responsible, helpful, forgiving, honest, loyal, mature love, a spiritual life, meaning in life,}} and \textit{\textbf{true friendship}} (c.f. \cite{schwartz1992universals}). Additionally, the theory has been widely accepted and adopted in several areas, including the social sciences, computer science, and software engineering \cite{perera2020study}.

Recent research on human values in SE has underscored the need for software companies to directly cater to issues of human values in their software development processes \cite{whittle2021case}, as the resulting software artefacts have a direct and indirect impact on end-users and society at large. Hussain et al. \cite{Hussain:2020} argue that the maturity levels of companies in addressing human values may very well depend on their awareness and overall organisational culture. They propose that incorporating human values should be done through the evolution of already established software practices, i.e., adapting existing processes to include human values considerations, e.g., the inclusion of values in the development of personas, rather than through a revolution of the field of SE. Winter et al. \cite{Winter:2018} proposed the values Q-sort method for measuring human values in SE. Applying the values Q-sort method to 12 software engineers shows 3 “software engineer” values prototypes. In a similar study, Shams et al. \cite{Shams:2021} applied the portrait values questionnaire (PVQ) to 193 Bangladeshi female farmers to elicit their values. Their study reported \textit{conformity} and \textit{security} as the most important value categories while \textit{power, hedonism,} and \textit{stimulation} were the least important for Bangladeshi female farmers.

Other studies have applied indirect approaches by using app reviews as a proxy for eliciting values requirements. Shams et al. \cite{shams2020society} analysed 1,522 reviews from 29 Bangladeshi agricultural apps to understand both the desired and missing values that should be addressed in the development of such apps. Furthermore, Obie et al. \cite{obie2021first}  introduced a dictionary-based natural language processing technique for detecting the violation of human values in app reviews. The result of their study showed that 26.5\% of the analyzed 22,119 app reviews contained perceived violations of human values by the end-users. In addition, \textit{benevolence} and \textit{self-direction} were the most violated categories while \textit{conformity} and \textit{tradition} were the least violated categories. 

As important steps towards addressing the violation of the values of mobile apps users in society, Obie et al. \cite{obie2021first} proposed the mining of values requirements from rich data sources, the alignment of values between stakeholders in SE projects, and the adoption of critical technical practice in mobile SE. A recent study further contends that careful consideration of domain context in the design and application of values instruments should be made during values requirements gathering, as the hierarchy of end-users values may vary depending on the end-users domain context \cite{obie2021does}.

Furthermore, Mougouei \cite{Mougouei2020engineering} proposed a framework for accounting for human values at the level of source code. This framework established a relationship between human values and Android APIs and includes the following aspects: annotating APIs with the relevant human values, inspecting source code to detect potential sources of values violations, and recommending fixes to mitigate the violations. Building on the work of Mougouei \cite{Mougouei2020engineering}, Li et al. \cite{li2021step} proffered 6 algorithms for detecting potential violation of values in 6 Android APIs. Their analysis applying these algorithms to 10,000 Android apps shows a correlation between the violation of human values and the presence of viruses in these apps.

As these studies have shown, the reflection of, support, and violation of human values may impact individual end-users and society as a whole. The research area of human values in SE is still in its early stages, and more work needs to be done. However, we present this work to further the discussion of human values in SE from the perspective of software developers as captured in their Stack Overflow posts.

\begin{table}
\small
\centering
\caption{Value categories and descriptions
\vspace{-3mm}
\cite{schwartz1992universals}}
\label{tab:valuescategories}
\begin{tabular}{lp{5cm}} 
\hline
\textbf{Value Category} & \textbf{Description}                                                                                         \\ 
\hline
Self-direction & Independent thought and action - choosing, creating, exploring                                                            \\ 
\hline
Stimulation    & Excitement, novelty, and challenge in life                                                                                \\ 
\hline
Hedonism       & Pleasure or sensuous gratification for oneself                                                                            \\ 
\hline
Achievement    & Personal success through demonstrating competence according to social standards                                           \\ 
\hline
Power          & Social status and prestige, control or dominance over people and resources                                                \\ 
\hline
Security~      & Safety, harmony, and stability of society, of relationships, and of self                                                  \\ 
\hline
Conformity     & Restraint of actions, inclinations, and impulses likely to upset or harm others and violate social expectations or norms  \\ 
\hline
Tradition      & Respect, commitment, and acceptance of the customs and ideas that one's culture or religion provides                      \\ 
\hline
Benevolence    & Preserving and enhancing the welfare of those with whom one is in frequent personal contact                               \\ 
\hline
Universalism   & Understanding, appreciation, tolerance, and protection for the welfare of all people and for nature                       \\
\hline
\end{tabular}
\end{table}

\subsection{Mining of Stack Overflow Posts}

Question and Answer (Q \& A) websites such as Stack Overflow are a rich source of information and provide insights into understanding developers’ behaviours, interactions, and viewpoints on specific topics amongst others \cite{Treude:2011}. Several studies have mined Stack Overflow posts and comments to shed light on key areas. For example, Novielli et al. \cite{Novielli:2014} focused on the social aspect of Stack Overflow and showed that the emotional lexicons in technical questions have an impact on the probability of obtaining satisfying responses to questions. Similarly, another study introduced multi-label classifiers for classifying the emotions encapsulated in Stack Overflow posts \cite{CABRERADIEGO:2020}.  
Wang et al. \cite{Wang:2013} analysed 100,000 questions from Stack Overflow to understand developer interactions on the platform. Key findings from their study show that developers are keen on contributing to the community and not just getting their questions answered; developers extend a helping hand to others whether or not their gesture is reciprocated. Similarly, another study found that being prompt and being the first to respond to questions helps quickly build a reputation on Stack Overflow \cite{Bosu:2013}.

Other studies on Stack Overflow have focused on specific technical topics, e.g., Bangash et al. \cite{Bangash:2019} analysed posts to understand what developers know about machine learning, while Bajaj et al. \cite{Bajaj:2014} reported on the challenges and misconceptions among web developers by mining posts related to HTML, CSS, and Javascript. Kavaler et al. \cite{Kavaler:2013} examined the relations between the Android marketplace usage of APIs to the occurrence of these APIs in Stack Overflow questions.

Some studies have relied on the rich dataset from Stack Overflow to develop tools for supporting software development. For instance, Ponzanelli et al. \cite{Ponzanelli:2014} introduced PROMPTER, an Eclipse plugin. Given a context in the IDE, PROMPTER automatically retrieves and analyses relevant discussions from Stack Overflow and then notifies the developer about available help. To support automatic source code documentation, Vassalo et al. \cite{Vassallo:2014} proposed CODES, a tool that extracts candidate method documentation from Stack Overflow discussions and creates Javadoc descriptions from it. 

The studies discussed above have been vital in understanding the various themes discussed on Stack Overflow and have also shown Stack Overflow as a rich data source for understanding these varied themes in discussions related to developers and the software development practice. We build on this body of knowledge by investigating developers' discussion from the important lens of human values and how it affects society.

\section{Data Collection}\label{sec:datacollection}

To understand human values violations in Stack Overflow and the possible reactions of its users to such violations, we first needed to identify a sufficient number of posts that have a values implication (e.g., violating a human value) (See Figure \ref{fig:motivationexp}). In the first data collection step, we executed a random SQL SELECT query on the Stack Overflow publicly available dataset\footnote{\url{https://tinyurl.com/4c74uz5n}} with Google BigQuery. This dataset, hosted through the Google Cloud Public Dataset Program \footnote{\url{https://tinyurl.com/y2x9rzch}}, is updated weekly by Stack Overflow and contains information about posts, comments, and voting, among other kinds of site activity. The SQL SELECT statement returned a random sample of 10000 comments and their corresponding posts, of which the first four authors (the analysts) individually manually analyzed the first 300. We imported the ID and content of these 300 comments and their corresponding posts into a spreadsheet. This spreadsheet was then shared between the analysts. The spreadsheet included 10 columns to enable the analysts to indicate which of the ten human value categories they judged were violated by the corresponding posts. The analysts then read each comment and its associated post (question or answer) to identify which of the ten value categories in Schwartz's theory (if any) were violated in the parent post according to the comment. 

Once the analysts finished this labeling process, they held several meetings and used a negotiated agreement method \cite{2013_campbell_coding, 1974_morrissey_sources} to resolve any disagreements and conflicts. Using the negotiated agreement method, all analysts collaboratively agreed on the label (coding) of an item under review. This approach is particularly useful for addressing reliability issues of codes when there are multiple categories as opposed to a binary category where a Cohen's Kappa measure would suffice. At the end of this step, we found a very low prevalence of comments (8 of 300 comments) that raised concerns about at least one of Schwartz's value categories.

In the second step, we designed a query to identify more posts that may potentially contain human values violations. To do so, we developed a list of regex keywords and phrases that are likely to be associated with human values violations concerns, such as “moral”, “ethical”, “human-cent”, and “society”. This list was designed based on our observations from the 300 comments and their corresponding posts labeled in the first step and consulting a dictionary of human values-related keywords and phrases developed in \cite{obie2021first} for identifying human values violations in app user reviews. This list was adjusted over several query runs, reducing to 21 regex keywords, and was used to design a SQL SELECT query to identify comments that contained one of the keywords in the list. The SQL SELECT query returned 10144 comments. Note that keywords have also been used in previous studies (e.g., \cite{zhang2019empirical, obie2021first}) to fine-tune data collection and minimize false positives. The list of 21 keywords and phrases is available in our replication package \cite{replication}.

In the final step, we randomly selected 2000 comments from 10144 comments, which is well above a significance level of 99\% and a significance interval of 3\%. We use these 2000 comments and their associated posts (1980 unique posts) to answer our research questions.

\section{Findings}\label{sec:findings}
\subsection{RQ1. Which types of human values are perceived to be violated in Stack Overflow posts?}\label{sec:RQ1}

\noindent\textbf{\underline{Approach}}. To identify human values violations in Stack Overflow, we qualitatively analyzed 2,000 comments and their associated posts (1,980 unique questions or answers) collected in the Data Collection section (See Section \ref{sec:datacollection}). We used Schwartz's theory of basic values \cite{schwartz2012overview}, specifically Schwartz's ten value categories (\textit{self-direction}, \textit{stimulation}, \textit{hedonism}, \textit{achievement}, \textit{power}, \textit{security}, \textit{conformity}, \textit{tradition}, \textit{benevolence}, \textit{universalism}), as a reference point to identify human values violations in Stack Overflow posts. This decision was made because Schwartz's theory of basic values is the most widely used and cited values model in social science and software engineering \cite{obie2021first, perera2020study, schwartz1992universals}. 
The 10 value categories, in turn, comprise 58 value items. In this study, we focus on the 10 value categories. Note that the treatment of the 58 individual value items is beyond the scope of this work. However, where appropriate, we refer to the relevant value items associated with the value categories to increase the clarity of the results.

We first created a spreadsheet and shared it with all authors. The spreadsheet included 15 columns. The first four columns recorded the Comment ID and the content of the 2000 comments and their associated post ID (question ID or answer ID) and the link to the posts. The next ten columns were the ten value categories. We also added a column called ``remark'' to allow the analysts to point out what they thought was important about a given comment/post.

The data analysis process was conducted in two steps. In the first step, the first author (the first analyst) followed an iterative process to label the first 1,000 comments and their associated posts. The first analyst selected approximately 100 comments and their associated posts and labeled them in each iteration. The reason behind investigating the associated posts was to thoroughly understand the context, meaning, and rationale behind comments. The first analyst was asked to indicate whether a comment discussing its associated post violated at least one human value. If so, they had to specify which of the ten value categories the given comment violated and put ``1'' in the corresponding columns in the spreadsheet. Comments could be labeled as violating more than one category of human value. After each iteration, three other authors (the validators) cross-checked the comments labeled in that iteration. In total, 400 comments (out of the 1,000 comments labeled by the first analyst) and their associated posts were cross-checked by the first validator, 400 by the second validator, and the rest by the third validator. This distribution was based on the availability of the validators. A negotiated agreement method \cite{2013_campbell_coding, 1974_morrissey_sources} was used to resolve conflicts and disagreements between the first analyst and the validators. The validators had extensive experience in human values and software engineering.

In the next step, the second analyst (the fifth author) labeled the rest of the comments (1,000 comments) and their corresponding posts. The second analyst conducted an iterative labeling process similar to the first analyst for this purpose. Then, the three validators in the previous step and the first analyst (acting as the fourth validator in this step) cross-checked the comments labeled by the second analyst in each iteration. In total, the first and second validators cross-checked 300 comments each, the third validator checked 100 comments, and the fourth validator cross-checked the rest. Similar to the previous step, the second analyst held several meetings with the validators to resolve disagreements and conflicts using the negotiated agreement method \cite{2013_campbell_coding, 1974_morrissey_sources}.

\noindent\textbf{\underline{Results}}. Our analysis of 2,000 Stack Overflow comments and their associated posts (1980 unique questions and answers) indicates that 315 comments (15.75\%) complained their corresponding posts (313 unique posts) violated human values (See our replication package \cite{replication}). Out of 10 Schwartz theory's value categories, we only found violations related to \textit{self-direction}, \textit{hedonism}, \textit{security}, \textit{conformity}, \textit{tradition}, \textit{benevolence}, and \textit{universalism} in the 315 comments. 
Our analysis did not find any violations regarding \textit{power}, \textit{achievement}, and \textit{stimulation}. 
The vast majority of the comments (270 out 315) include concerns that explained their corresponding post violated the value of \textit{hedonism}. An example of a comment highlighting the violation of the value of \textit{hedonism} (value item of \textit{pleasure}) is:

\faThumbsDown{} \textit{``Interesting approach and a possible solution, but \textbf{not very user-friendly. The user won't see the proper day values when he is moving the slider} (unless I write another JS function to take care of that).''} \textsc{\textbf{Comment ID: 35155704}} 


The value of \textit{benevolence} is the second most frequently violated value (reported in 41 comments). For example, a Stack Overflow user raised the issue of the violation of \textit{benevolence} (value item of \textit{responsibility}) and criticized the irresponsibility of another user because their proposed solution does not care about the sensitivity of information but only about the cost of the solution.

\faThumbsDown{} \textit{``What I meant was that when it comes to sensitive information, your attitude of ``SSL is too expensive'' is \textbf{unethical and irresponsible}. When you have sensitive information in your hands, you must do everything you can to secure it. You say that you don't see its worth in ``common business cases'', but ``common business cases'' often involve sensitive information (addresses, phone numbers, email messages, trade secrets, etc). A business has much to lose in a breach of data integrity.''} \textsc{\textbf{Comment ID: 4919033}}

We found only 10 comments in which Stack Overflow users complained that the proposed solutions violate \textit{security}.

\faThumbsDown{} \textit{``Still, \textbf{passwords are private to the person} that fills it in during registration. Not encrypting them \textbf{is not very ethical}, but I guess that’s another subject to discuss.''} \textsc{\textbf{Comment ID: 3334429}}


Eight comments were found complaining about violating \textit{conformity}. In the following example, a user criticized the poster as their approach to terminate an app (proposed in a question, \textsc{\textbf{Question ID: 3318806)}} violates Apple's User Interface standards and guidelines. 

\faThumbsDown{} \textit{``I think your app may get rejected if you terminate it within the app (unless due to unrecoverable error/fault handling). \textbf{Apple doesn't like you to mess with their user experience, and pressing the Home button to exit/suspend an app is a big part of that user experience}.''} \textsc{\textbf{Comment ID: 3442304}}

For \textit{self-direction} and \textit{universalism}, we found five and four examples of human values violation, respectively, and two for \textit{tradition}. For example, the comment below is mapped to violation of the value of \textit{self-direction} as its associated post violates the \textit{freedom} and \textit{independence} of end-users (i.e., \textit{freedom} and \textit{independence} are two value items in the value category of \textit{self-direction}).    

\faThumbsDown{} \textit{``Not to mention it strikes me as ethically dicey at best to \textbf{grab a user’s location without their permission}.''} \textsc{\textbf{Comment ID: 16973544}}


\subsection{RQ2. How quick are commenters to raise concerns over human values violations?}
\noindent\textbf{\underline{Approach}}. To answer this RQ, we extracted the date of the 315 comments about values violations and the date of their corresponding posts and calculated the time difference. To further understand how quickly comments about values violations are raised, we compared the position of a post's comment that voices out about values violation with that of all of its associated comments.     

\begin{figure*}
    \centering
    \includegraphics[width=0.80\linewidth]{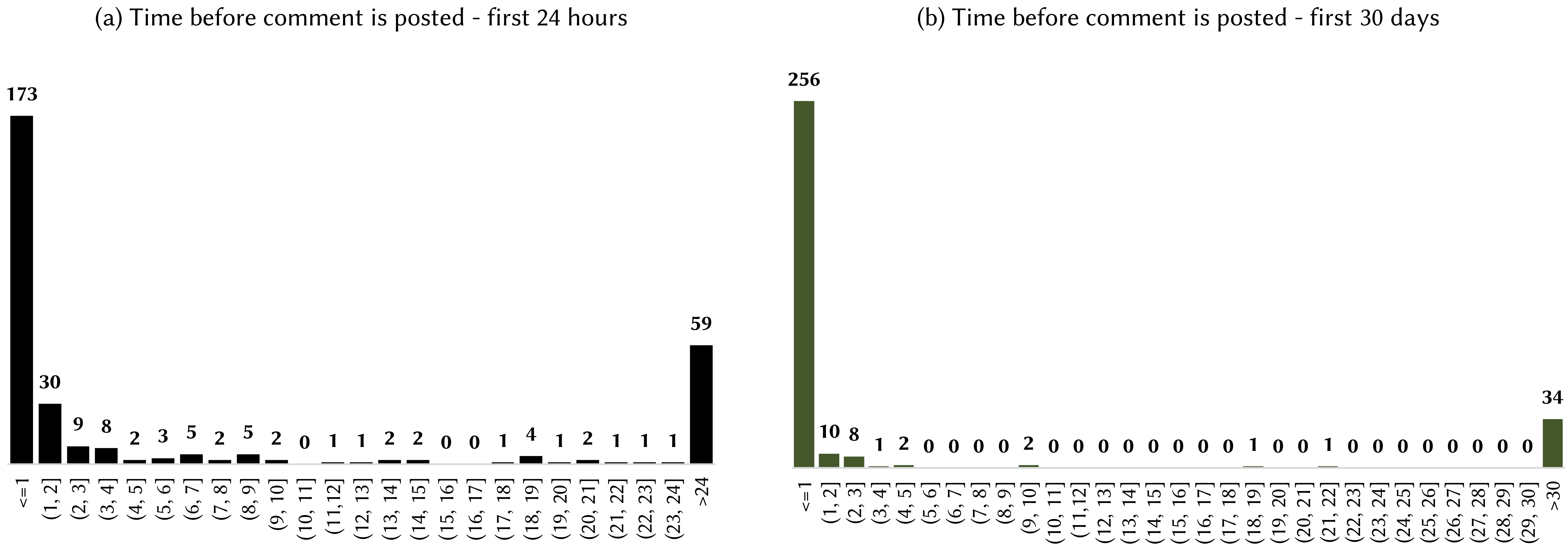}
    \vspace{-3mm}
    \caption{The time taken for concerns about violated human values to be raised (N=315)}
    \label{fig:timeinterval}
\end{figure*}

\begin{figure}
    \centering
    \includegraphics[width=0.75\linewidth]{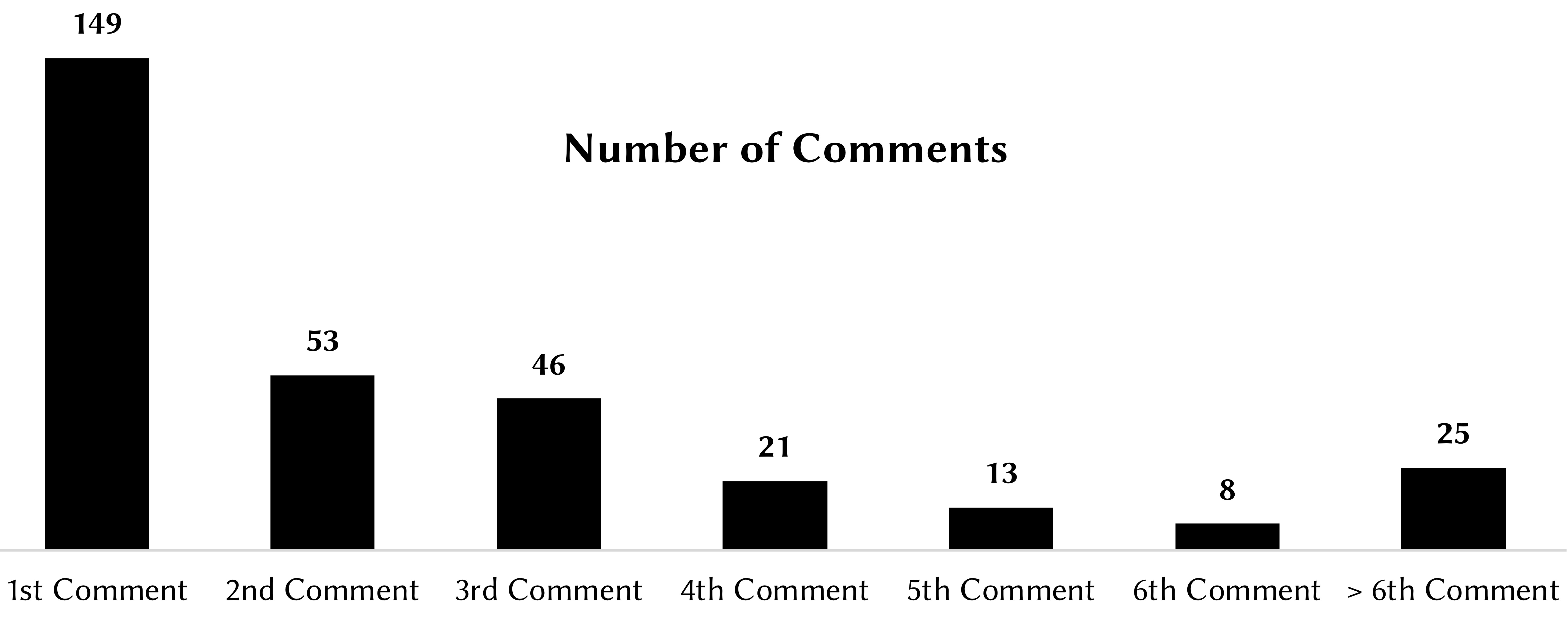}
    \vspace{-3mm}
    \caption{Positions of values violations comments (N=315)}
    \label{fig:commentposition}
\end{figure}

\noindent\textbf{\underline{Results}}. Figure \ref{fig:timeinterval} depicts the time differences in hours and days. It is shown that almost 55\% of comments (173 out of 315) about values violations were received less than one hour after the corresponding posts were made. Our analysis shows that only 59 comments were raised 24 hours after the date of their corresponding posts. As shown in Figure \ref{fig:commentposition}, comments that raised concerns about human values mostly were the first or second comments (202 out of 315 comments, 64.12\%) of their corresponding questions or answers. Only 25 comments about values violation appeared after the sixth comment.

\subsection{RQ3. Are posts accused of violating human values downvoted by Stack Overflow users?}


\noindent\textbf{\underline{Approach}}. We developed an SQL Query to count the downvotes that were cast on the 313 unique posts associated with the 315 human values violation comments following the creation of those comments.

\noindent\textbf{\underline{Results}}. We found that Stack Overflow users did not vote down most posts accused of violating human values violation. In fact, out of 313 posts with human values violation comments, only 74 (23.64\%) posts were downvoted after comments complaining about human values violations were raised. Out of these 74 downvoted posts, most were downvoted once (46 posts), followed by twice (10 posts) and three times (10 posts). The rest were downvoted four times (7 posts) and five times (1 post). 

For example, a user asked the question (\textsc{\textbf{Question ID: 6854611}}):
\textit{``How to return value when AJAX request is succeeded.''} The accepted answer suggested setting parameter \textsf{async} to \textsf{false} to address the problem. However, a Stack Overflow user criticized this solution because it negatively impacts user experience. The accepted answer received three downvotes after the comment was made.

\faThumbsDown{} \textit{``Async: false will \textbf{LOCK UP THE BROWSER} for the duration of the ajax call. \textbf{This is often a horrible user experience.} The good solution requires refactoring the code to use a callback or a function call from the success/error functions to continue execution of the process and pass on the result of the ajax call.''} \textsc{\textbf{Comment ID: 8151749}}

In another example, a user sought a solution to automatically turn on the user phone's GPS as the given android app launches (\textsc{\textbf{Question ID: 17723347}}). Another user indicated that such solutions violate \textit{user privacy}. Such solutions can also violate the \textit{self-direction} value (specifically, the value item of \textit{independence}) as it denies the \textit{independence} of the user to choose and set their security and privacy settings. This question got downvoted once.

\faThumbsDown{} \textit{``You can't and you shouldn't - it's \textbf{ethically questionable to override something the user has set} that has security/privacy considerations.''} \textsc{\textbf{Comment ID: 25833312}}

\subsection{RQ4. How does the original poster react to concerns of potential human values violations?}


\subsubsection{RQ4.1 Does the original poster respond to concerns about potential human values violations in further comments? If so, how?} \hspace{1cm}

\noindent\textbf{\underline{Approach}}. To understand if and how the original poster responds to a comment raising concerns about human values violation in a post, we conducted a qualitative study on all comments made by the original posters after the date of the 315 human value violations comments. We found that the original posters added 288 comments after the 315 human value violations comments. The first author applied the open coding technique \cite{glaser1968discovery} on all the 288 posters' comments to decide which one(s) were added to respond to the reported human value violations and categorized the original posters' responses (e.g., denying the post violates a human value). To reflect the distribution of original posters' reaction on Stack Overflow, we limited tagging to at most one response comment one per human values violation comment; this way, each original poster's response to their associated human values concern would be counted once, if, e.g., the poster created multiple comments after the concern was raised. In the next step, the outcomes of the open coding process (i.e., identified codes and categories) were shared with the second author for review. Then, the first and second authors held several Zoom meetings to discuss disagreements and inconsistencies and reached a consensus on the final list of codes and categories. 

\noindent\textbf{\underline{Results}}. 
Our analysis shows that out of the 288 comments added by original posters after a comment citing a human values violation, 140 did not address the violation concern. The remaining 148 comments addressed the human values violation concern directly, but the exact nature of the responses varied widely. The qualitative analysis of these 148 comments using open coding led to their classification as one of six response types.

The first four response types - namely ``Acknowledging their violation'', ``Adding redemptive detail'', ``Proposing an alternative'', and ``Asking for clarification'' - were all responses, in which the original poster was receptive to commenter's concern about human values violations in their post. Receptive responses totaled 105 of the 148 relevant responses.

\textit{\textbf{Acknowledging their violation (n=44)}}: In these responses, the original poster accepted the commenter's concern wholly as described. This was often accompanied by an attempt to mitigate the violation, either by adapting their original proposed solution or by abandoning their approach altogether.

\textit{``@David, I think you are right. The user experience of editing a Rich TextBox within a DataGridView is rather bad, so I followed your hint on providing a separate edit mask. However, I'm also stuck here. See: http://stackoverflow.com/questions/10224556/how-to-edit-a-dataset-in-a-new-form.''} \textsc{\textbf{Comment ID: 13134970}} replied to \textsc{\textbf{Comment ID: 13112316}}

\textit{\textbf{Adding redemptive detail (n=40)}}: In this type of response, the original poster added further information about the context and/or requirements of their solution to explain why their post does not violate the human value claimed by the commenter.

\textit{``I need to do it. It is a special app that is meant to change desktop background, get system settings, disable icons, auto-hide task bar etc. It is used in one of the computer stores to display computer sale sticker directly on the screen.''} \textsc{\textbf{Comment ID: 9302848}} replied to \textsc{\textbf{Comment ID: 9302833}}

\textit{\textbf{Proposing an alternative (n=13)}}: In these responses, the original poster responded with an alternative solution in an attempt to mitigate the violation while still achieving their objective.

\textit{``In that case then you can queue users selection and updated it one by one, or if it is possible to stop the current operation, stop it and start the new fresh thread.''} \textsc{\textbf{Comment ID: 28900232}} replied to \textsc{\textbf{Comment ID: 28887851}}

\textit{\textbf{Asking for clarification (n=8)}}: In these responses, the original poster asked for further information about the commenter's concern. These responses reflected a willingness to engage with the concern raised and to further inquire as to whether their solution is violating human values.

\textit{``@JPReddy: Sorry, I didn't understand your problem, the changing must affect only the ComboBox control, so it must not affect any other cell in this column or other column, so can you explain more.''} \textsc{\textbf{Comment ID: 4324670}} replied to \textsc{\textbf{Comment ID: 4324160}}

The remaining two response types - ``Denying'' and ``Conceding and pressing on with the issue'' - were dismissive, rather than receptive, to the human values concerns raised earlier in the comment thread.

\textbf{\textit{Denying (n=28)}}: In these responses, the original poster denied that their post violated human values at all.

 \textit{``@Stephen: No. `Expensive' isn't `unethical'. It's just free market economy. The school didn't have to give the job to that guy. It CHOSE to do so. They could always look for alternatives and choose the cheaper offer.''} 
\textsc{\textbf{Comment ID: 1749489}} replied to \textsc{\textbf{Comment ID: 1748956}}

\textit{\textbf{Conceding and pressing on with the issue (n=15)}}: In these responses, the original poster wholly accepted that there was a human values violation in their post but asserted that they were nonetheless going to persist with their approach.

 \textit{``Preventing me from viewing a site simply because I'm on desktop because they want to feed me more ads and JS spam isn't particularly ethical either. Oh well..''} \textsc{\textbf{Comment ID: 59368548}} replied to \textsc{\textbf{Comment ID: 59304750}}

\subsubsection{RQ4.2 Does the original poster modify their original post in light of concerns about human values violations? If so, how?} \hspace{1cm}        

\noindent\textbf{\underline{Approach}}. Some original posters may go beyond a comment response to a values violation claim, and may modify their original post itself in light of the claim. We first collected posts associated with the 315 comments which cited human values violations. Then we checked how many of them were modified after those comments were added. We found that 103 posts had at least one modification made on or after the time that the human values violation comment was created. Next, the first author manually checked all activities carried out on the 103 posts after the time of the 315 human value violations comments to understand which ones were related to the reported values violations.  

\noindent\textbf{\underline{Results}}. Out of 103 posts modified by its poster after the reported human values violations, only 14 posts were edited by the original poster in response to the claim about human values violations. The rest (89 posts) were modified for other reasons. These 14 posts can be grouped into two categories. Nine posts were edited to mitigate human value violations. For example, an original poster changed the source code in their post (\textsc{\textbf{Post ID: 15135545}}) after receiving criticism from a user who indicated the proposed solution \textit{``will cause the GUI to completely freeze, which is not a good user experience''} \textsc{\textbf{Comment ID: 21307680}}. 

In the other category of edit, the original poster edited their post to reinforce that no human values violation existed - this was true of five posts. For example, although an original poster edited their post (\textsc{\textbf{Post ID: 16660109)}}, the solution in the post was not modified, and the original poster only added more information to clarify why there was no user experience issue with their solution.

\section{Discussion and Implications}\label{sec:discussion}
\subsection{Prevalence of Hedonism Violations}
The distribution of the types of values violations in our dataset was largely dominated by \textit{hedonism} violations. The Schwartz model defines the \textit{hedonism} value category as “\textit{pleasure or sensuous gratification for oneself}”, and associated with it are the value items \textit{pleasure}, \textit{enjoying life}, and \textit{self-indulgent} \cite{schwartz2012overview}. As such, any comment in the dataset which voiced concern about a post’s negative impact on users’ pleasure and experience would be tagged as a \textit{hedonism} violation. 
Moreover, the value of a programming solution is largely determined by how the user feels when interacting with it, and user experience requirements are typically included within the technical requirements for software projects. Thus, when commenters voice concerns about a violation of \textit{hedonism}, they may be addressing the technical requirements of the software rather than incidental, unintended consequences on human values. As a result, any objections on technical user experience grounds to a solution would be considered a complaint about a violation of the \textit{hedonism} value category.

\textbf{\underline{Implications:}} This finding raises the need to re-think the paradigms used for analyzing human values discourse in software contexts. When the concern for human values falls directly in line with software technical requirements --as in the case of \textit{hedonism}-- it is important to account for the distinction of this dynamic from other cases where human values come at the expense of satisfying and delivering technical requirements. We claim that values conforming to technical goals is precisely the objective of values discourse: the more the software community voices their opinions about the importance of human values, the more those values become a fixed feature of software quality. Regardless, the dynamics between human values and technical requirements should be directly addressed in further research in these fields.

\subsection{Reactions to Posts Violating Values} 
We investigated the reactions to posts that violated human values from the perspectives of Stack Overflow commenters, users, and the original posters. We found that commenters are quick in terms of raising concerns about human values violations. However, our study shows that Stack Overflow users did not downvote most posts (76.35\%) accused of violating values. This may not be surprising as Stack Overflow users are encouraged to use upvotes and downvotes to report on the quality of the information in posts. A downvote on a question means \textit{“this question does not show any research effort; it is unclear or not useful”}, and, on an answer, \textit{“this answer is not useful”} \cite{stackoverflow_downvoting}. As such, downvotes are not intended to reflect perceived violations of human values per se. A question post that is well-crafted and thought-out, while containing blatant human values violations, could avoid any downvotes, and so too with an answer post. Conversely, any downvotes on a post containing a human values violation may have nothing to do with the values violations in question but rather with the quality and usefulness of the post.

Nevertheless, it is useful to observe Stack Overflow users’ actual behavior in casting downvotes. Our results indicate that downvotes are, to some extent, cast in the wake of a human values violation being voiced in the comments. Indeed, it is possible that users are employing downvotes as a way to cast disapproval in light of perceived values violations, despite official site guidelines; how well users comply with site guidelines in their site activity is a major topic of discussion on the Meta Stack Exchange site \cite{stackoverflow_whendownvoting}.

\textbf{\underline{Implications: }} Further research is needed to measure the significance of our  findings on downvotes in human values contexts against general patterns of downvoting on Stack Overflow posts. This would shed light both on users’ reactions to perceived values violations on Stack Overflow and their regard for site voting guidelines in general.

\subsection{Reactions to the Accusation of Human Values Violation in Posts} 

We observed while the original posters usually acknowledge commenters who criticize their posts, they tended to downplay the severity of the issue, either by minimising the impact of the violation, or by justifying their decision in spite of the severity of the issue. While being receptive is a good characteristic, we emphasize that the original posters need to mitigate values violations in their posts actively. This can potentially avoid the possible risks that such posts may have on the end-user and society. There is also the need for (Stack Overflow) developers to consider human values and the potential violation of these values in the technical solutions that they proffer in these platforms. We also recommend the consideration of other currently less investigated human values such as \textit{achievement, tradition}, and  \textit{conformity} (as categorised by Schwartz \cite{schwartz1992universals}) beyond the well-researched values of \textit{privacy} and \textit{security}. 

\textbf{\underline{Implications: }}  Investigating the nature of values discourse on Stack Overflow could find relationships between the language and format used to voice concerns about value violations and the types of reactions they evoke from authors of the software in question. This would allow researchers to understand what makes users more or less receptive to criticism when it comes to the implications of their work on human values.

\subsection{Towards an Automated Tool}
Given our study is an exploratory study, we chose a manual approach to identify comments containing concerns about human values violations and categorize the response comments from the original posters to such values violations comments. This limited our sample size, as manual methods become more time-consuming as the amount of data scales up. 

\textbf{\underline{Implications: }} Automated approaches using machine learning and natural language processing methods should be developed to detect comments raising concerns about human values violations. In this study, we mainly used the contents of comments under posts to recognize posts violating values. AI-based techniques could leverage other features (e.g., votes, response comments by the original poster, the reputation score of the commenter and poster) to detect such posts and the types of values violated by these posts. Such automated methods will inform Stack Overflow users of a possible values implications of a post and let them decide if they want to use solutions proposed in the post in their software systems \cite{nurwidyantoro2021human,alomar2021finding}.


\section{Threats to Validity}\label{sec:threatstovalidity}
\textbf{External Validity} refers to what extent our findings can be generalized to other contexts \cite{wohlin2012experimentation}. This study collected and analyzed a random sample of only 2000 comments and their associated posts (questions or answers) from a dataset of comments and their associated posts described in Section \ref{sec:datacollection}. So, we acknowledge that our findings may not be generalized to all posts and comments in Stack Overflow and other question and answer websites such as Reddit\footnote{\url{https://www.reddit.com/}} and Gitter\footnote{\url{https://gitter.im/}}. Further research needs to be conducted to explore how posts in other question and answer websites violate human values and how their users react to such violations.

\textbf{Internal Validity} is defined as threats that may have impacted our findings \cite{wohlin2012experimentation}.
The random selection of the 2000 comments from a dataset of 10144 comments and the qualitative analysis processes conducted for RQ1 and RQ4 may have threatened our findings. First, our decision to build a dataset of 10144 comments from millions of comments in Stack Overflow was motivated to reduce the number of false-positive comments as much as possible. Furthermore, it was not possible for us to manually analyze all 10144 comments. Hence, we analyzed a random sample of 2000 comments and their corresponding posts. Therefore, we may have missed some important types of value violations because of our dataset in Section \ref{sec:datacollection} and the random selection of the 2000 comments from the dataset. Furthermore, some of the phrases used to build the dataset are particularly hedonism-related terms, which may have allowed the \textit{hedonism} category to be over-represented in our dataset, resulting in a skew towards hedonism comments in RQ1.

The qualitative analysis processes to answer RQ1 and RQ4 might be subjective and error-prone. In RQ1, we employed two approaches to reduce these issues. First, the assigned analysts were asked to analyze the data iteratively (in each iteration, only 100 comments and their corresponding posts were analyzed by the analyst). Second, once each set of 100 comments and their associated posts were analyzed, the validators cross-checked these comments labeled by each analyst. Furthermore, several meetings were organized between the analysts and validators to resolve disagreements and conflicts using the negotiated agreement method \cite{2013_campbell_coding, 1974_morrissey_sources}. In RQ4, the second author checked all categories and their corresponding codes, and any disagreements and conflicts were resolved through meetings. In both RQ1 and RQ4, there were comments and their associated posts that made it difficult for us to precisely identify the type of human values violations (RQ1) or the type of the poster's responses (RQ4). In such cases, we labeled the comment as a non-human values comment (RQ1) or as an irrelevant response (RQ4) to avoid possible risks and mistakes. Hence, we can be reasonably confident that our findings are credible with minimum mislabelled comments.

\textbf{Construct Validity}. In RQ1, our decision to use ten value categories in the Schwartz theory may have introduced two threats. First, other values models such as Rokeach’s Value
Survey \cite{rokeach1973nature} and List of Values \cite{kahle1988using} with different types and numbers of values could be used instead of the Schwartz theory. While none of them have been developed for software engineering, the Schwartz theory is widely used in software engineering (e.g., \cite{perera2020study,nurwidyantoro2021human}). Furthermore, the definition of human values might have been vague for the analysts and validators. So, they might have struggled to map Stack Overflow comments to human values. To mitigate this threat, apart from reading seminal papers \cite{schwartz1992universals, schwartz2012overview} on the Schwartz theory, we consulted the previous research that leveraged (the definition of) human values in the software engineering context, e.g., app reviews \cite{obie2021first,shams2020society}, GitHub issue discussions \cite{nurwidyantoro2021human}, and source codes \cite{Mougouei2020engineering}. Finally, the quantitative measures used in RQ2, RQ3, and R4 to determine the characteristics of comments citing values violations or reactions to posts being accused of values violations would not capture all aspects of these comments and posts. Future research is encouraged to further characterize these comments and posts. 
\section{Conclusion and Future Work}\label{sec:conclusions}

In this work, we conducted an exploratory study investigating the potential violation of human values in Stack Overflow. Adopting the widely accepted Schwartz model of basic human values, we analyzed 2,000 Stack Overflow comments and their associated posts (1980 unique questions or answers) to identify posts and comments containing perceived human values violations, the categories of the values violated, and the reactions of Stack Overflow users to concerns related to these violations. Our results show that 315 (out of 2,000) comments raised issues concerning the violation of 7 out of the 10 value categories in the Schwartz model. We find that Stack Overflow commenters react quickly to issues of values violations; 203 (out of 315) comments raising the concerns of values violations were made less than 2 hours after the corresponding posts. Also, most posts (76.35\%) accused of human values violation did not get downvoted at all. Furthermore, only 148 of the original posters responded to the concerns of values violations made by other commenters in follow-up comments of their own.

In the future, we plan to build upon our exploratory study by diving deeper into specific value categories and their associated value items to understand the different factors that cause their violations. In addition, due to the limitations of a manual approach to categorizing values and their violations, we plan to build machine learning models to automate this process. 




\section*{Acknowledgments}
Support for this work from ARC Laureate Program FL190100035 and Discovery Project DP200100020 is gratefully acknowledged.

\bibliographystyle{ACM-Reference-Format}
\bibliography{main}

\end{document}